\documentclass[conference]{IEEEtran}
\IEEEoverridecommandlockouts
\usepackage{cite}
\usepackage{amsmath,amssymb,amsfonts}
\usepackage{algorithmic}
\usepackage{graphicx}
\usepackage{textcomp}
\usepackage{xcolor}
\usepackage[nolist]{acronym}
\usepackage{array}
\usepackage{makecell}
\def\BibTeX{{\rm B\kern-.05em{\sc i\kern-.025em b}\kern-.08em
    T\kern-.1667em\lower.7ex\hbox{E}\kern-.125emX}}

\begin{document}




\newacro{US}[U.S.]{United States}
\newacro{FBI}[FBI]{Federal Bureau of Investigations}
\newacro{CIA}[CIA]{Central Intelligence Agency}



\begin{acronym}[XXXXXXX] 

\acro{API}{application programming interface}
\acro{AV}{authentication vector}
\acro{DOD}{Department of Defense}
\acro{DoS}{denial of service}
\acro{NPS}{Naval Postgraduate School}
\acro{TCP}{Transmission Control Protocol}
\acro{USN}{\ac{US} Navy}
\acro{JADC2}{Joint All-Domain Command and Control}
\acro{DMO}{Distributed Maritime Operations}
\acro{EAB}{expeditionary advance base}
\acro{EABO}{Expeditionary Advance Base Operations}
\acro{3G}{3rd generation wireless}
\acro{4G}{4th generation wireless}
\acro{5G}{5th generation wireless}
\acro{5GS}{\ac{5G} system}
\acro{UE}{user equipment}
\acro{IoT}{internet of things}
\acro{RAN}{radio access network}
\acro{DU}{distributed unit}
\acro{CU}{centralized unit}
\acro{ORAN}{Open Radio Access Networks}
\acro{ZTA}{Zero-Trust Architecture}
\acro{TDL}{tactical data link}
\acro{NF}{network function}
\acro{VNF}{virtual \ac{NF}}
\acro{NIDS}{network intrusion detection system}
\acro{vNIDS}{virtual \ac{NIDS}}
\acro{vFW}{virtual firewall}
\acro{5GC}{\ac{5G} core network}
\acro{3GPP}{3G Partnership Project}
\acro{URLLC}{ultra-reliable and low-latency communication}
\acro{mMTC}{massive machine type communication}
\acro{AR}{augmented reality}
\acro{VR}{virtual reality}
\acro{AI}{artificial intelligence}
\acro{ML}{machine learning}
\acro{Mb}{megabits}
\acro{Gb}{gigabits}
\acro{Mb/s}{\ac{Mb} per second}
\acro{Gb/s}{\ac{Gb} per second}
\acro{RIC}{\ac{RAN} intelligent controller}
\acro{KPI}{key performance indicators}
\acro{gNB}{\ac{5G} base station}
\acro{RRC}{radio resource control}
\acro{PDCP}{packet data convergence protocol}
\acro{SDAP}{service data adaptation protocol}
\acro{NTN}{non-terrestrial network}
\acro{AMF}{Access and Mobility Management Function}
\acro{AUSF}{Authentication Server Function}
\acro{NSSF}{Network Slice Selection Function}
\acro{SMF}{Session Management Function}
\acro{UPF}{User Plane Function}
\acro{PCF}{Policy Control Function}
\acro{UDM}{Unified Data Management function}
\acro{SBA}{service based architecture}
\acro{PDU}{protocol data unit}
\acro{DN}{data network}
\acro{LDN}{local data network}
\acro{MEC}{Mobile edge computing}
\acro{SDN}{software defined networking}
\acro{UDR}{Unified Data Repository}
\acro{SEAF}{Security Anchor Function}
\acro{SUPI}{Subscriber Permanent Identifier}
\acro{SUCI}{Subscriber Concealed Identifier}
\acro{NSSAI}{network slice selection assistance information}
\acro{5G-AKA}{\ac{5G} authentication and key agreement}
\acro{USIM}{universal subscriber identity module}
\acro{NAS}{Non-Access Stratum}
\acro{GUTI}{globally unique temporary identifier}
\acro{QoS}{quality or service}
\acro{TTL}{time-to-live}

\end{acronym}

\title{A Case for Enabling Delegation of 5G Core Decisions to the RAN}

\author{\IEEEauthorblockN{1\textsuperscript{st} Lucas Vancina}
\IEEEauthorblockA{\textit{Computer Science Department} \\
\textit{Naval Postgraduate School}\\
Monterey, California, USA \\
lucas.vancina@nps.edu}
\and
\IEEEauthorblockN{2\textsuperscript{nd} Geoffrey Xie}
\IEEEauthorblockA{\textit{Computer Science Department} \\
\textit{Naval Postgraduate School}\\
Monterey, California, USA \\
xie@nps.edu}

}

\maketitle

\begin{abstract}


Under conventional 5G system design, the authentication and continuous monitoring of \ac{UE} demands a reliable backhaul connection between the radio access network (RAN) and the core network functions (AMF, AUSF, UDM, etc.). This is not a given, especially in disaster response and military operations. We propose that, in these scenarios, decisions made by core functions can be effectively delegated to the RAN by leveraging the RAN's computing resources and the micro-service programmability  
of the O-RAN system architecture. This paper presents several concrete designs of core-RAN decision delegation, including caching of core decisions and replicating some of the core decision logic. Each design has revealed interesting performance and security trade-offs that warrant further investigation.

\end{abstract}


\section{Introduction}
\label{sec:intro}



\ac{5G} networks are designed not only to provide elevated performance and security, but also to meet diverse and ever-evolving requirements from both users and operators. To achieve such a level of flexibility, the 5G control plane forms a logically centralized decision element within the \ac{5GC}, which relies heavily on virtualized core network control functions to handle \ac{UE} registration (the AMF function), authentication (the AUSF, SEAF, and UDM functions), resource allocation (the NSSF function), security monitoring (AMF and SMF), and mobility management (AMF). In this architecture, another distinct component is the \ac{RAN}, also referred to as a {gNodeB} or a base station in the literature, which provides physical-layer radio access to \ac{UE}s and performs corresponding link layer functions. 

This design makes an implicit  assumption that the network connection between the \ac{RAN} and the \ac{5GC} functions (i.e., the ``backhaul" links) will be reliable. We observe that the assumption may not always hold. To illustrate, consider the following three scenarios: 
\begin{itemize}

    \item {\bf Disaster response operations.} Large-scale disasters often severely damage the network infrastructure of the affected areas. A critical requirement in disaster relief is to establish, \emph{as quickly as possible}, some level of data network connectivity for the population and the responders. Backhaul links between rapidly installed \ac{RAN}s and \ac{5GC}, typically over ad hoc wireless connections with limited bandwidth, may be prone to overload and outage. 

    \item {\bf Military operations.} Integration of 5G into forward-deployed military tactical networks has similar challenges of rapid set-up and ad hoc wireless connections. Additionally, tactical networks commonly face austere and adversarial networking conditions that exacerbate the problem. 
    
\begin{figure}[t]
\centering
\includegraphics[scale=0.31,trim=0 5 0 10, clip]{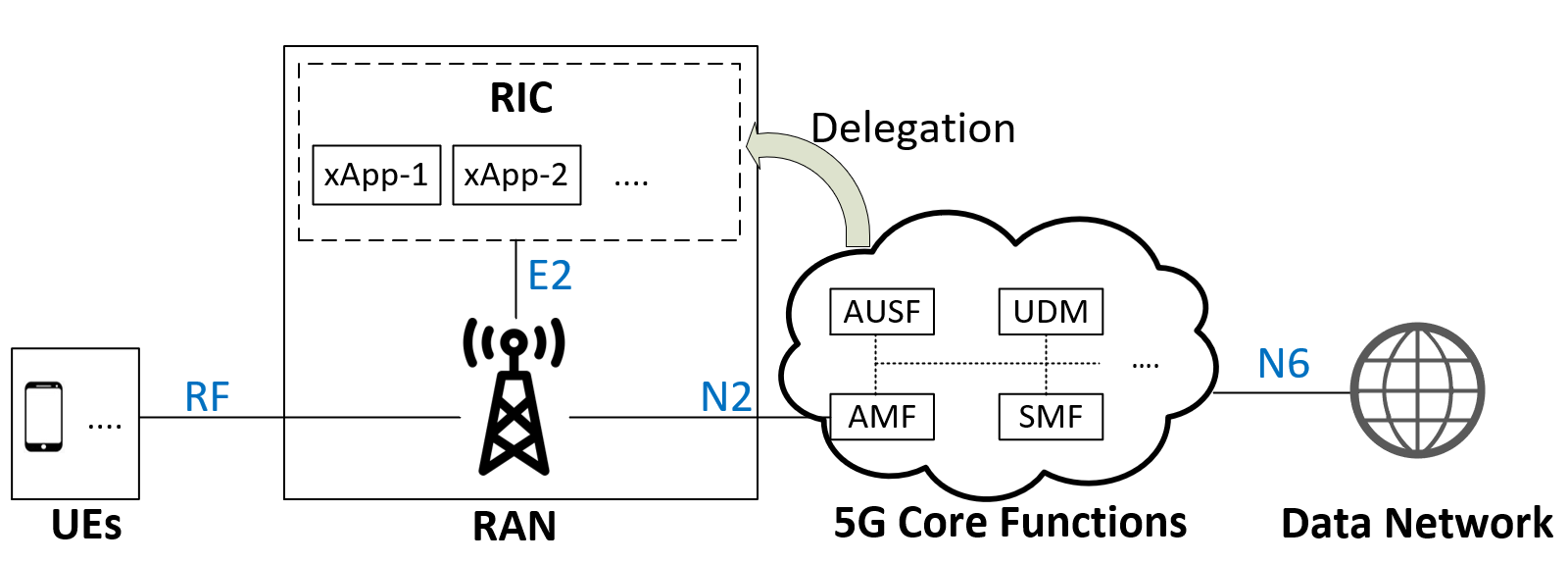}
\caption{Concept of Core-RAN decision delegation. Custom-designed xApps in the RAN will act on behalf of the core decision functions when required to allow \acp{UE} to connect and access designated resources.}
\label{fig:concept}
\end{figure}


    \item {\bf Flash crowd congestion} Consider a 5G operator that provides edge computing services to mobile users including pedestrians and vehicles in a large city. When an excessive number of service requests arrive at an edge server at the same time in a ``flash crowd'' situation, the backhaul link for the corresponding RAN may be overloaded, causing severe degradation of user experience. 

    
\end{itemize}



In these scenarios, it is highly desirable that an affected RAN is able to maintain some levels of service for local UEs even if its backhaul links have become unavailable. For example, the RAN should continue to allow the UEs to communicate with each other and access online resources nearby such as edge computing servers. We observe that one approach to realizing such functionality is to delegate some of the core control decisions to the RAN, as illustrated in Figure~\ref{fig:concept}. In particular, 
ensuring that security control decisions can continue to be carried out with a degree of independence from the \ac{5GC} is important because all authentication currently involves the core.


Equally important, the Open RAN (ORAN) system architecture~\cite{oran-alliance}, the emerging architecture of choice for 5G RAN implementations, provides a software programming environment. Specifically, as illustrated in Figure~\ref{fig:concept}, a RAN Intelligent Controller (RIC) is able to orchestrate a custom collection of micro-service modules called xApps. This software extensibility may be ideally suited for deploying custom solutions for delegating core decisions to the RANs.  

In this paper, we explore three concrete points of the design space for core-RAN delegation solutions: (i) collocated \ac{5GC} deployment,  (ii) caching of core decisions, (iii) combination of caching and partial replication of control logic. We detail the unique challenges and trade-offs of each design in terms of performance and security. 
Recognizing that it is impossible to conduct a meaningful study for each of the vast collections of 5G core control functions at the same time, we have chosen to focus on delegating security control decisions in this work. 

The rest of the paper is organized as follows. Section 2 provides a tutorial of relevant 5G concepts and discusses related work. Section 3 presents the three designs of core-RAN decision delegation. Additional use cases and possible extensions are discussed in Section 4. Finally, Section 5 concludes the paper.


\section{Background and Related Work}

\label{sec:related}

The first two parts of this section provide background information about the \ac{5GC} decision process and \ac{RAN} capabilities, respectively. The third part of the section is a brief review of the related work.


\subsection{\ac{5G} Core-Centric Decision Process}
\label{sec:core_process}


%

\begin{figure}[tb]
\centering
\includegraphics[scale=0.65,trim=0 5 0 10, clip]{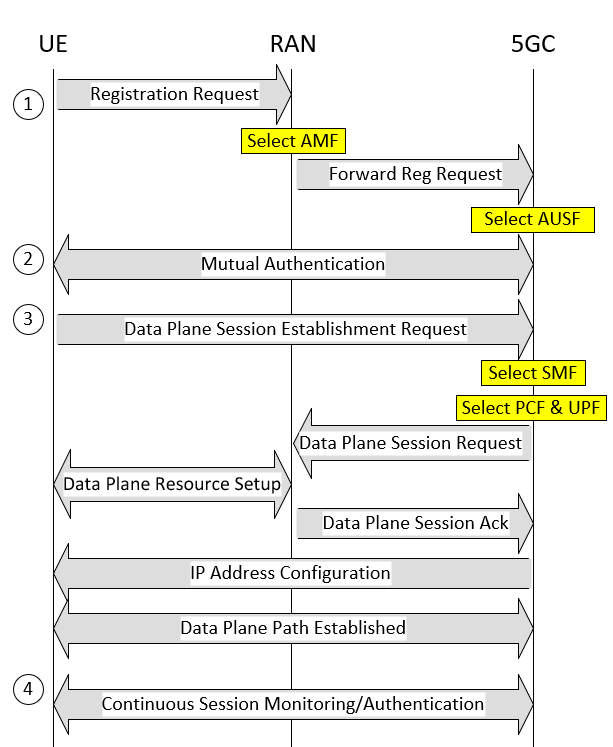}
\caption{Illustration of the four main steps that a \ac{UE} must go through involving \ac{5GC} network control functions (NFs) before the \ac{UE} is able to send and receive data over a 5G network. \ac{NF} selections are made from a set of \ac{NF} instances based on \ac{UE} subscription and service requirements, performance consideration, and other policy.}
\label{fig:process}
\end{figure}

The \ac{3GPP} 5G standard  
defines a network control decision process that heavily depends on the virtual control functions located  in the \ac{5GC}.  We illustrate the four major steps of this decision process in 
Figure~\ref{fig:process}. 

At the first step, upon the UE's connection request, the \ac{RAN} selects an \ac{AMF} instance based on the identity and context information provided by the \ac{UE} and forwards the registration request to that \ac{AMF} within the \ac{5GC}. The \ac{AMF} then selects an \ac{AUSF} at the \ac{UE}'s home network and sends it an authentication request containing the \ac{UE} identifier \cite{lei_5g_2019}.

At the second step, the \ac{AUSF} gets the UE's subscription and authentication data from the \ac{UDM} and initiates a mutual authentication with the \ac{UE}~\cite{basin_5G-security_18}.

At the third step, the newly authenticated \ac{UE} sends a session establishment request to the \ac{AMF}. The \ac{AMF} coordinates the selection of a \ac{SMF} to manage the session, which in turn selects a \ac{PCF} to provide relevant policy information and one or more \acp{UPF} to perform the actual data plane control. Once the control plane \acp{NF} are established, a data plane session request is sent back to the \ac{RAN} and the \ac{RAN} allocates radio and compute resources to handle the specified session requirements and communicates that to the  \ac{UE}. The \ac{UE} then receives an internet protocol address from the \ac{SMF} and is able to send data securely to the \ac{UPF}, completing the session establishment \cite{lei_5g_2019}.

The fourth step is the ongoing process of monitoring the session for security and \ac{QoS} and reauthenticating as needed based on security policies.

\subsection{{RAN} Intelligent Controller and {xApps} Capabilities}


\ac{ORAN}~\cite{oran-alliance} is a modular design that allows operators to combine software and hardware from any vendor that conforms to the required interface standards. 
As another defining characteristic,  \ac{ORAN} is a \emph{software-defined system}, much like the \ac{5GC}, with strong support for software extensibility. Its programming environment consists of the following two major components.  

\subsubsection{RAN Intelligent Controller}
In \ac{ORAN} the bulk of the control plane functionality is orchestrated in a software entity called the \ac{RIC}. The \ac{RIC} acts like a brain for the \ac{RAN}, gathering data from the data plane components through standardized interfaces and acting on the data and policy to provide instructions to optimize the performance of the \ac{RAN}. The \ac{RIC} is divided into the near-real-time \ac{RIC} to perform time critical control (with a sub-second latency requirement) and the non-real-time \ac{RIC} to handle other non time critical tasks~\cite{polese_understanding_2023}.

The \ac{RIC} supports software extensibility \emph{explicitly with a service orchestration framework} to deploy self-contained software modules (referred to as micro-services). Custom micro-services can be added and registered on the \ac{RIC} to be invoked when specific network conditions arise~\cite{polese_understanding_2023}.

\subsubsection{xApps}

The micro-services in the near-real-time category are called xApps. 
\ac{ORAN} features xApp software development kits including frameworks that allow vendors as well as open source developers to easily customize features that can run on any \ac{ORAN} compliant \ac{RIC}, primarily using the C++, Go, and Python programming languages \cite{noauthor_overview_nodate}. Custom xApps have been developed to provide a wide range of functionality, from machine learning driven performance optimization and analytics \cite{qazzaz_machine_2024}, to security functions~\cite{wen_fine-grained_2022,haohuang_wen_5g-spector_2024}, to traffic steering and slice management \cite{johnson_nexran_2021}.

Standardized APIs make it possible for custom xApps to interact seamlessly with the \ac{ORAN} control and data plane entities. Different combinations of xApps from a variety of vendors can work together to support performance and security goals.  Because xApps run in the near-real-time control loop of the \ac{RIC}, they are expected to respond in between 10 milliseconds and 1 second.\cite{polese_understanding_2023} 


\nocite{kholidy_toward_2022}

\subsection{Related work}

We review related work in three aspects. First, as mentioned earlier, several prior studies~\cite{qazzaz_machine_2024,haohuang_wen_5g-spector_2024,johnson_nexran_2021} have successfully utilized custom xApps to enhance the RAN performance and security. However, they do not address the limitations of the 5G core-centric decision process. 

Second, in a broad sense, our work aims to increase the autonomy of edge systems so that they can continue to function at some capacity after losing access to the core infrastructure. There has been a long line of prior work with similar goals, from exploring hastily formed networks~\cite{rawat_towards_2015} 
to developing edge computing technology~\cite{pham_survey_20}. 

Finally, we briefly note two more 5G security investigations for their close relevance to this work. The first study~\cite{basin_5G-security_18} is a formal and detailed analysis of the 5G authentication protocol suite, which we have used as a sanity check for our understanding of the 5G authentication process. The second study~\cite{raavi_core-attack_21} reveals that a 5G core network may be susceptible to DoS attacks, which can be considered another scenario to motivate our research.  







\section{Design of Core Decision Delegation}
\label{sec:design}


This section presents three design alternatives to achieving core-RAN decision delegation and, as a result, reducing dependence on the network backhaul.
\begin{itemize}
  \item {\bf Collocated \ac{5GC} deployment.} Deploy a full 5GC instance that is physically collocated with the \ac{RAN}.

  \item {\bf Decision caching.} Cache \ac{5GC} control decisions so that future repeat requests can be handled by the \ac{RAN} according to the cached decisions.

  \item {\bf Decision logic replication.} Replicate selected \ac{5GC} control decision logic and caching of associated core state to allow the \ac{RAN} to act as a proxy for some core network functions.
    
\end{itemize}  

Given our focus on security control, we start the section with an analysis of 5G security key management. 

\subsection{Analysis of \ac{5G} Key Management}
A critical aspect of any network security system is cryptographic key management. The \ac{5G-AKA} protocol is the standard protocol used for a \ac{5G} network and a \ac{UE} to mutually authenticate and coordinate cipher suites. \ac{5G-AKA} implements a hierarchy of keys that are all some level of derivation from shared secret $K$, a symmetric key that is installed on the \ac{UE}'s \ac{USIM} and in  the \ac{UDR} of the \ac{UE}'s home network~\cite{3gpp_technical_specification_group_ts33501_2024}.

\begin{figure}[h]
\centering
\includegraphics[scale=0.5,trim=0 5 0 10, clip]{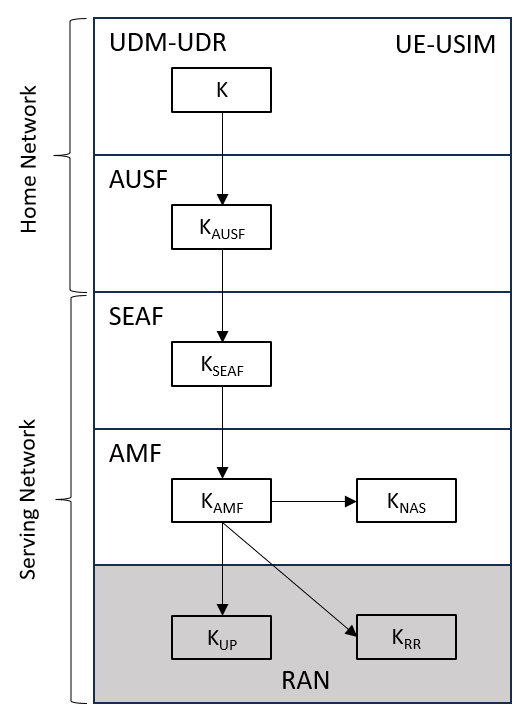}
\caption{A visualization of the 5G-AKA key derivation hierarchy. The \ac{UE}'s \ac{USIM} and its home network's \ac{5GC} \ac{UDR} both possess shared secret $K$. The \ac{5GC} \acp{NF} and \ac{UE} are able to use $K$ to generate subsequent keys.}
\label{fig:hierarchy}
\end{figure}

Figure \ref{fig:hierarchy} is a simplified version of the \ac{5G} key derivation hierarchy diagram from reference~\cite{lei_5g_2019} and shows a sequence of three major key derivations performed by \acp{NF}. 
{\bf (i)} When the \ac{AUSF} receives an authentication request, it requests authentication materials from the \ac{UDM} and receives $K_{AUSF}$ derived from $K$, along with an authentication challenge token and the corresponding expected response salted by $K$. 
This tuple of a new key, authentication challenge, and expected response is called the \ac{AV}.
{\bf (ii)} Next, the \ac{AUSF} derives $K_{SEAF}$, a semi-permanent authentication key, and sends it to the \ac{SEAF} as part of the \ac{AV}. The \ac{SEAF} sends the authentication challenge to the subscriber \ac{UE} which sends back a response, generated using $K$. This response is checked against the expected response from the \ac{UDM} to confirm that the \ac{UE} possesses the correct $K$, thus proving its identity. {\bf (iii)} Lastly, the \ac{SEAF} generates from $K_{SEAF}$ the control plane encryption keys ($K_{AMF}$ and $K_{NAS}$) and data plan encryption keys ($K_{UP}$ and $K_{RRC}$) specific for that \ac{UE}.

Note the division between the serving network, which is the \ac{5GC} that the \ac{UE} is requesting services from, and the home network, which is the network that holds $K$ for that \ac{UE} as well as its subscription information. These can both be the same network, but, when roaming, a \ac{UE} connects to a serving network that does not share that \ac{UE}'s $K$, in which case the serving network must refer the authentication request to the \ac{UE}'s specified home network. The home network will then produce an \ac{AV} for the serving network as discussed, allowing the \ac{SEAF} to act as an authentication proxy for the home network.


We make the following observations:

\begin{enumerate}
    \item Authentication of a new \ac{UE} requires access to the shared secret, $K$. Without $K$ there is no practical way to generate an \ac{AV} that can confirm the identity of the device.
    \item When the serving network is different from the home network (i.e. when roaming), the serving network must establish a \ac{SEAF} and request authentication keys from the home network's \ac{AUSF} before it can authenticate the new \ac{UE}. However, $K_{SEAF}$ allows a serving network entity to provide a range of authentication services for that \ac{UE} in the future.
    \item The cascading structure for key derivation means that, as long as a \ac{NF} holds a key higher in the hierarchy than the one needed, then that \ac{NF} should be able to derive the necessary lower-level key.
\end{enumerate}


This concludes the analysis of 5G key management. The next three subsections present the three design alternatives for core-RAN decision delegation.

\subsection{Design 1: Collocated \ac{5GC} Deployment}
The first design alternative is to physically collocate the \ac{5GC} and the \ac{RAN}. Because all of the core \acp{NF} are software defined and able to be virtualized~\cite{3gpp_technical_specification_group_ts_2023}, the \ac{5GC} can be run on the same physical machine or in the same data center as the \ac{RAN}, eliminating the need for using backhaul to access core control functions. 

\subsubsection{Performance}
Attaching a fully functional \ac{5GC} at every base station might improve latency without a major system redesign, and thus a viable solution for a single-RAN 5G network,  but these ``Distributed Cores" will still need to be federated to a ``Super Core" when deploying RANs across multiple geographical locations. Federation can lead to a significant amount of overhead in network communications and added complexity for state synchronization to handle roaming \ac{UE}s.  The memory and processing requirements associated with running a full \ac{5GC} instance may not be feasible without costly hardware upgrades to base stations.


\subsubsection{Security}
A collocated core would certainly yield significant improvements in availability because it would provide the RAN with local access to core NFs. Physical security would be a significant risk, however, since the symmetric keys ($K$) would be stored on hardware close to the network edge instead of a secure data center. This would counteract the layered security approach to 5G system design that seeks to protect the master keys deep inside the core \cite{3gpp_technical_specification_group_ts33501_2024}.

\subsection{Design 2: Decision Caching}
The second design alternative is to deploy a custom xApp at the \ac{RAN} to maintain a cache for each request from eligible \ac{UE}s and the corresponding responses from the \ac{5GC} \acp{NF}. The \ac{RAN} then responds directly to repeat requests when cached decisions are available, without having to forward the requests across the backhaul link. 
The set of \ac{UE}s eligible for decision caching can be pre-configured according to network policy.

Table \ref{tab:decision_cache} shows the type of records that the Core Decision Cache would maintain to allow for low latency processing of network registration and authentication requests. Each record must contain a unique identifier for the originating UE (this can be hashed to provide additional privacy), necessary keys for ciphering/deciphering of both control and user plane messages, tables that map a specific request type to the corresponding  sets of data plane and control plane specific messages to be distributed by the \ac{RIC}, and a \ac{TTL}. 

Proactive decision caching would also enable an ``Express Mode" for registration, allowing specific \ac{UE}s to register and re-authenticate extremely quickly. This would improve efficiency for \ac{UE}s that only transmit data periodically, such as nodes in a sensor network or unmanned system, because they would not have to maintain a connection in between transmissions. 
Again, the set of \ac{UE}s eligible for the express mode can be pre-configured at the RAN.

\subsubsection{Performance}
Caching is a proven technique for reducing latency in accessing network resources. By caching decisions from the core with their associated actions, it is reasonable to expect a significant performance boost when there are repeated accesses from a UE. The magnitude of the improvement will depend on the situation and the level of sophistication of the algorithms used to determine when and how long to cache core decisions.

\subsubsection{Security}
Removing the necessity to store $K$ on the base station hardware would be a distinct security improvement over the previous design alternative. If a malicious actor were able to steal $K_{SEAF}$, they should not be able to deduce any past or future version of that key \cite{3gpp_technical_specification_group_ts33501_2024}, so the impact would be much lower than if $K$ were compromised. On the other hand, further research should be done to devise an effective system to prevent replay attacks. \ac{5G-AKA} includes randomness and sequence numbers to prevent replay based authentication loopholes \cite{basin_5G-security_18}, but caching credentials might introduce potential for a combination of man-in-the-middle and spoofing exploits.


\begin{table*}
    \vspace{.05in}
    \centering
    \begin{tabular}{|p{2.5cm}|p{2cm}|p{5cm}|p{5cm}|p{1cm}|}
        \hline
        \multicolumn{1}{|c|}{\bf \ac{UE} Cached ID} & \multicolumn{1}{c|}{\bf Cached Keys} & \multicolumn{1}{c|}{\bf Cached Control Plane Decisions} & \multicolumn{1}{c|}{\bf Cached Data Plane Decisions} & \multicolumn{1}{c|}{\bf TTL} \\
        \hline
        \hline
        e.g. $SHA_{256}$(SUCI) & e.g. $K_{SEAF}$ & e.g. Registration Req : \{Access controls, ...\} & e.g. Registration Req : \{Traffic steering, ...\} & e.g. 60:00 m\\
        \hline
    \end{tabular}
    \caption{Illustration of Core Decision Cache for \acp{UE}}
    \label{tab:decision_cache}
\end{table*}

\begin{table*}
    \centering
    \begin{tabular}{|p{2.5cm}|p{2cm}|p{5cm}|p{5cm}|p{1cm}|}
        \hline
        \multicolumn{1}{|c|}{\bf \ac{UE} Cached ID} & \multicolumn{1}{c|}{\bf Cached Keys} & \multicolumn{1}{c|}{\bf Cached Control Plane State} & \multicolumn{1}{c|}{\bf Cached Data Plane State} & \multicolumn{1}{c|}{\bf TTL} \\
        \hline
        \hline
        e.g. $SHA_{256}$(SUCI) & e.g. $K_{SEAF}$ & e.g. Registration Req : \{Access controls,...\};\newline Subscription plan, authorized services,\newline preferences, ... & e.g. Registration Req : \{Traffic steering,...\};\newline Resource allocation, \ac{QoS} requirements, ... & e.g. 60:00 m\\
        \hline
    \end{tabular}
    \caption{Illustration of Core State Cache for \acp{UE}}
    \label{tab:state_cache}
\end{table*}

\subsection{Design 3: Decision Logic Replication}
\begin{figure}[tb]
\centering
\includegraphics[scale=0.6,trim=0 5 0 10, clip]{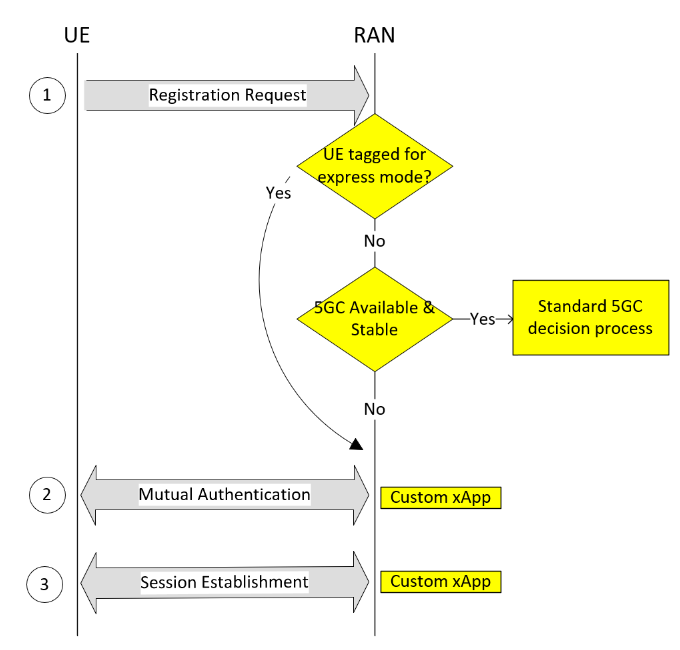}
\caption{Illustration of decision logic replication. Custom xApps 
act as a proxy for the core \acp{NF} when required. The major steps are the same as in Figure~\ref{fig:process}. However, at step 1, this design (i) supports the ``express mode" for selected UEs with cached decisions, and (ii) is able to take over authentication and session establishment for other eligible UEs with cached core state when the backhaul is congested or unavailable. 
}
\label{fig:logic_rep}
\end{figure}

The final alternative we devised is to deploy one or more xApps to implement specific portions of core \ac{NF} logic within the \ac{RAN} as illustrated in Figure~\ref{fig:logic_rep}. The design uses a generalized version of the Core Decision Cache, called the Core State Cache, which is shown in Table~\ref{tab:state_cache}. It caches the same data as the Core Decision Cache and additionally, core state information about the \ac{UE}, such as subscription and policy data. The primary steps are the same as the standard call flow in Figure~\ref{fig:process}, but with additional logic for scenarios where connectivity to the \ac{5GC} is not available and stable:

\begin{itemize}
    \item \textbf{Step 1}: When the UE submits a registration request, the \ac{RIC} checks if that UE is tagged for express processing. If so, then the request is immediately referred to an xApp for automated session establishment in accordance with the cached decisions. 
    \item \textbf{Step 2}: The health and available bandwidth of the backhaul link is assessed. If there is a strong connection, then registration will continue normally.
    \item \textbf{Step 3}:  Otherwise, the registration process will be diverted to a chain of custom xApps which use the Core State Cache data to authenticate and handle the UE's request.
\end{itemize}


For example, an autonomous mobile surveillance drone that provides early warning of forest fires does not need to maintain a constant network connection while executing its patrols, but it must be able to quickly  contact edge computing servers with alert messages if it identifies a fire, or just to log sensor readings into a database or get instructions from its command and control server. With this design, the RAN will have the option to trigger express mode, using proactively cached core decision data to ensure fast registration and delivery of alert messages. For less urgent requests, like those for the database, the RAN will also be able to access the drone's subscription and policy data from the Core State Cache to handle its requests even if the backhaul link is temporarily disrupted.



\subsubsection{Performance}

This design would allow \ac{UE} requests to be handled via the standard method when backhaul connectivity is stable, and at the same time provide a method to keep local communications and edge computing services available during congestion or outage of the backhaul. It also supports the express registration mode to further increase the performance of eligible UEs. While the design will introduce additional resource consumption at the RAN, we believe the consumption would be within the resource allocation of the ORAN framework. 

\subsubsection{Security}
Similar security benefits and weaknesses to those for Design 2 (Decision Caching) are applicable here. Upon closer look, however, this design provides potential for new solutions to mitigate security attacks because it taps more into the programmablility of the xApps. We describe an example solution for mitigating denial of service attacks  in Section~\ref{sec:solution-attacks}.

\section{Discussion}
\label{sec:discussion}

\subsection{Additional Use Cases}

\subsubsection{Industrial IoT}
Enabling automation, remote monitoring, and command and control of industrial \ac{IoT} are key use cases for \ac{5G}. Industrial \ac{IoT} systems often require particularly low latency and strict security controls \cite{peto_5g_2020}. For industries such as oil and gas production and delivery, ad hoc wireless connections are also common because of limited networking infrastructure in some operating areas. Enabling RAN autonomy will be beneficial for these IoT systems.  

\subsubsection{Non Terrestrial Networks}


5G \acp{NTN} are gaining popularity and satellites are expected to be called upon to service more and more mobile devices \cite{rinaldi_non-terrestrial_2020}. In some 5G NTN designs\cite{campana_NTN_23}, satellites function as the RAN -- referred to as non-terrestrial RANs -- while the core is hosted in the terrestrial network. A key challenge with satellites, however, is their limited capacity and susceptibility to atmospheric interference. Mitigating interference from atmospheric conditions and managing signal bandwidth are key considerations in satellite communications \cite{rinaldi_non-terrestrial_2020}. 
Consequently, the 5G backhaul capacity and reliability will likely become a performance limiting factor for such NTNs for the foreseeable future.  Delegating core functionality to the \ac{RAN} has the potential to mitigate this limitation.




\subsection{Extensions}


\subsubsection{Probationary Authentication}
Consider a disaster response or military operation where mobile devices from many organizations converge on a particular area and require network services on shared infrastructure. It would take significant time and effort for the network administrators to migrate all of the necessary subscription data over in order to conduct authentication as described in Section~\ref{sec:core_process}. Degraded or intermittent backhaul connectivity would exacerbate the problem and limit network availability for many of the devices. 

By delegating decision capabilities to the \ac{RAN}, however, a system of ``Probationary Authentication" could be implemented, allowing a roaming \ac{UE} to access a specially monitored network slice without full authentication. This would allow the \ac{UE} to access non-sensitive network services in a controlled environment. Once the UE's home network is contacted and full authentication performed, that UE would then be transferred to a regular network slice and its state would be cached for future reference. The availability of basic communications is valuable in rapid network deployment scenarios where timely actions are often critical.

\subsubsection{Denial of Service Attack Mitigation}
\label{sec:solution-attacks}

Handling registration and authentication requests at the \ac{RAN} offers the capability to filter out malicious traffic before it ever reaches the \ac{5GC}. A \ac{DoS} attack, similar to the flash crowd scenario discussed in Section \ref{sec:intro}, has the potential to saturate the available bandwidth of the backhaul link and cripple the network. Empowering the \ac{RAN} to filter and triage registration requests 
would ensure that suspicious requests are dropped before they can consume backhaul bandwidth.

\subsubsection{Zero Trust}
\ac{ZTA} is a recently advocated approach to network security that focuses on protection of data rather than network boundaries \cite{rose_zero_2020}. 5G network security would greatly benefit from the application of \ac{ZTA} principles due to its numerous software defined entities and open interfaces \cite{kholidy_toward_2022}.  However, implementing \ac{ZTA} adds more burden of authentication and monitoring of related network functions as \ac{UE}s are required to be continuously authenticated when accessing network resources. Allowing the \ac{RAN} to handle some repeat authentications would help mitigate \ac{ZTA}-induced congestion on the backhaul.


\section{Conclusion}

This paper presented three possible designs for delegating 5G core control decisions to custom software running on the RANs. The potential benefits as well as the technical challenges associated with such delegation were also explored in detail. We believe that the motivating use cases along with the design options that we put forward make a compelling case for pursuing this direction of research. For future work, we plan to build system prototypes to evaluate the performance of these designs in realistic settings and propose practical solutions to technical obstacles.


\bibliographystyle{IEEEtran}
\bibliography{references}

\end{document}